\documentclass[aps,prl,twocolumn,groupedaddress]{revtex4}
\usepackage{graphicx} 
\usepackage{amssymb,amsmath}

\begin{document}
\title{Band Structure and Quantum Conductance of Nanostructures
from Maximally-Localized Wannier Functions:\\
The Case of Functionalized Carbon Nanotubes}
\author{Young-Su  Lee,$^{1}$   Marco  Buongiorno  Nardelli,$^{2}$  and
Nicola Marzari$^{1}$}
\affiliation{$^{1}$Department of Materials Science and Engineering and
Institute  for Soldier  Nanotechnologies,  Massachusetts Institute  of
Technology, Cambridge, MA 02139, USA 
\\ $^{2}$Center for High Performance Simulations (CHiPS) and Department of
Physics,  North  Carolina State  University,  Raleigh, NC 27695, USA 
\\  and CCS-CSM, Oak Ridge National  Laboratory, Oak Ridge,
TN 37831, USA}

\date{\today}

\begin{abstract}
We  have combined large-scale, $\Gamma$-point electronic-structure
calculations with the maximally-localized Wannier functions approach
to calculate efficiently the band structure and the
quantum conductance of complex systems containing thousands of atoms
while maintaining full first-principles accuracy.  
We have applied this approach to study covalent functionalizations in 
metallic single-walled carbon nanotubes.
We find  that the band structure around the Fermi
energy is  much less dependent on  the chemical nature  of the ligands
than on the $sp^3$ functionalization pattern disrupting the
conjugation  network. Common aryl functionalizations are more stable
when paired with saturating hydrogens; even when paired, they still
act as strong scattering centers that degrade the ballistic conductance
of the nanotubes already at low degrees of coverage.
\end{abstract}
\maketitle

\begin{figure}[t]
\includegraphics[width=2.8 in]{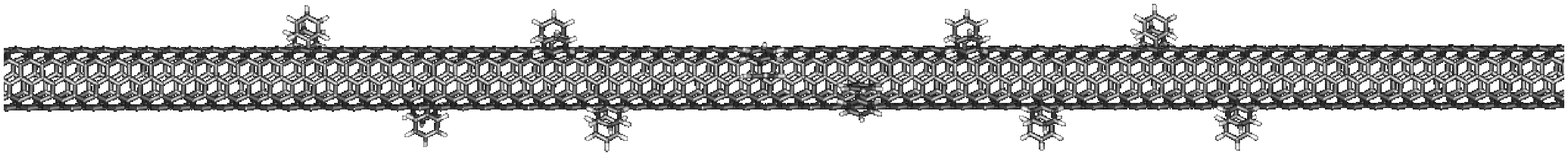}
\centering
\includegraphics[width=2.8 in]{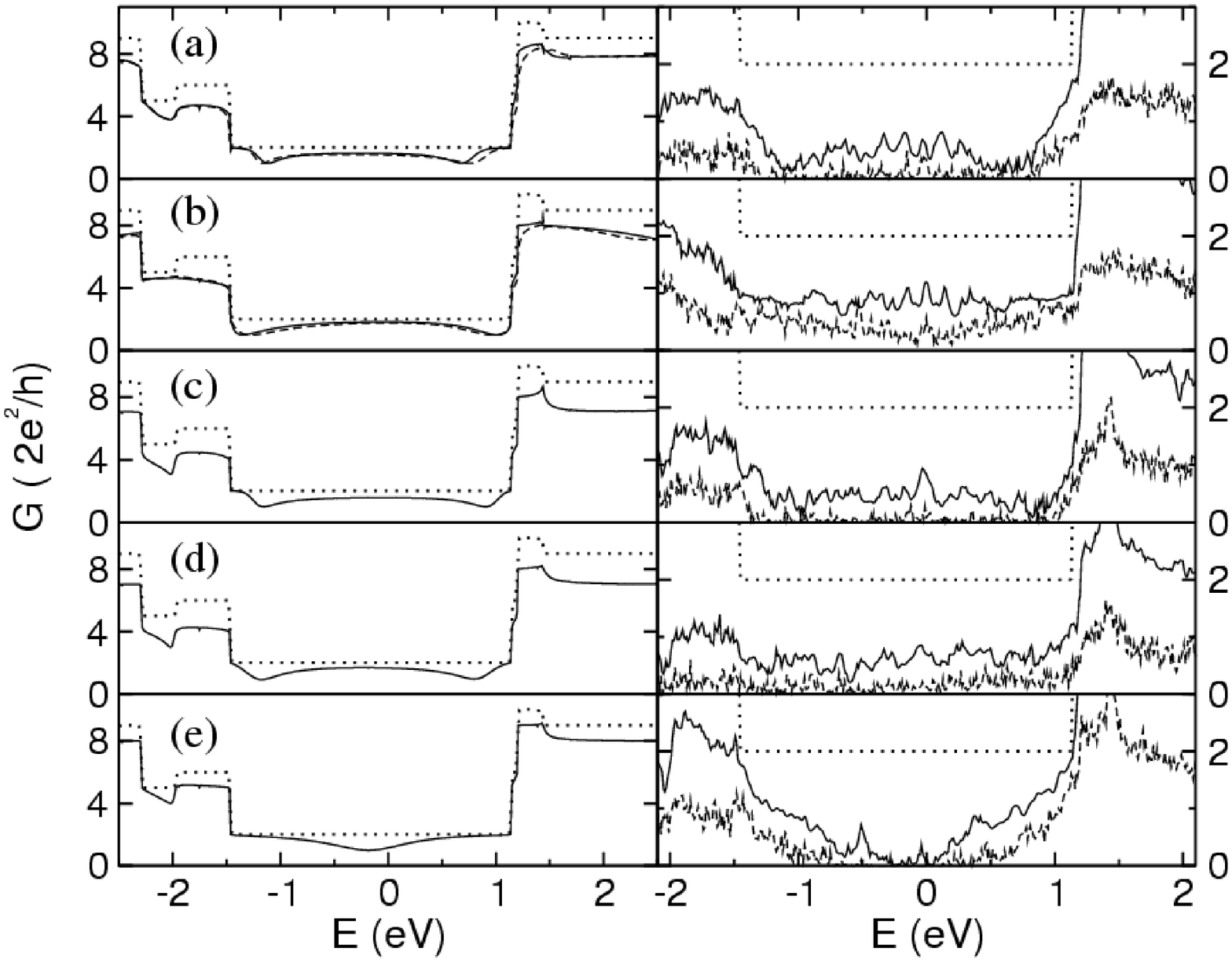}
\caption{
Top: infinite metallic (5,5) CNT functionalized in its central region
by an array of phenyl pairs.
Left panel: quantum  conductance of an infinite (5,5) CNT  with one 
isolated pair of ligands in positions $\alpha$ (a), $\beta$ (b), 
$\gamma$ (c), $\delta$ (d), or with a single ligand attached (e)
(dashed line: ligands are hydrogen atoms; solid line: model calculation;
dotted line: pristine CNT).  Right panel:
random distributions of pairs of ligands  or of single ligands
(arranged as in the left panel), for the case
of 10 defects per 1000 carbons (solid line) or 
30 per 3000 (dashed line), averaged over 5 random configurations.}
\label{fig:scatter}
\end{figure}

Carbon nanotubes (CNTs) are some of the most promising
materials in micro- and nano-electronics, whose applications could range from
transistors to interconnects to memories \cite{Avouris}. 
In the present effort to engineer and tune
the electronic properties of CNTs, 
organic functionalizations of the sidewalls provide one of the most encouraging
avenues \cite{FuncRev,Strano03}, 
ideally combining the exceptional electronic properties of 
CNTs with the diversity of organic chemistry.
In  this  Letter,  we study the effects of covalent functionalizations on
single-walled CNTs using a novel  approach  to  calculate  the
electronic structure and quantum conductance, based on our
idea of constructing well-localized and transferable 
orbitals as ideal building blocks for the electronic structure of complex
nanostructures \cite{Marzari97,Souza02,Calzolari04,Thygesen04}. 
These maximally-localized Wannier functions (MLWFs) provide 
a most compact and most accurate local representation of the
electronic structure of a solid or a molecule, and
are used to construct the full Hamiltonian
and to derive the band structure and the Landauer conductance of complex systems 
containing at least thousands of atoms, preserving first-principles accuracy 
and with linear-scaling computational costs. 
\\
Pristine, single-walled CNTs are ideal one-dimensional quantum wires with quasi-ballistic
electron transport for tens of nanometers (at low bias) even at room temperature \cite{Javey04,Park04}.
Structural or electronic defects and inelastic excitations
are then some of the most relevant
factors which limit conductance.
Covalent functionalizations
will alter conduction manifolds and doping levels, thus affecting
significantly the electronic transport properties of CNTs \cite{Kamaras03}. 
First-principles approaches
are especially suited to characterize such
effects, in the controlled environment of quantum-mechanical simulations.
Here, we  use density-functional theory  in the
GGA-PBE approximation \cite{Perdew96}, using state-of-the-art
solid-state approaches \cite{ESPRESSO}.
Explicit first-principles calculations of systems containing thousands of atoms
are rarely possible \cite{Wang97}, and the delocalized nature of the Bloch orbitals is not well-suited
to describe the transport properties of aperiodic systems;
these obstacles are overcome by the approach described in the following.
\\
{\it Disentanglement and  Localization:} 
Our method starts from the construction of an optimal MLWFs basis set.
In a large supercell, where the sampling of the Brillouin zone (BZ) is limited
to the $\Gamma$-point,
MLWFs can be obtained  by  a unitary  transformation 
of  the occupied ground-state eigenfunctions  $\{\vert\psi_m\rangle\}$:
$\vert w_n\rangle=\sum_{m=1}^{N_{occ}}
\,U_{mn}\,\vert\psi_m\rangle$,  where $U$ is chosen to
minimize the sum of the spread of all $\vert w_n\rangle$ around their
centers \cite{Marzari97,Silvestrelli98}. 
This is not a viable approach for metallic systems, or even for
insulators where the occupied subspace can significantly change character
as we move away from $\Gamma$, even if
for the supercell at hand
$\Gamma$-sampling might be adequate to calculate
the energy and charge density of the system with the necessary accuracy.
Our goal is instead to obtain, from a combination of
the occupied and some of the unoccupied eigenfunctions at $\Gamma$,
a compact and localized set of orbitals that describes 
a group of bands of interest (i.e., manifold) in the whole BZ
of the supercell - similar in spirit to the $\bf k\cdot p$ method \cite{Scandolo00},
but with a different approach and improved outcome.
This optimal set is obtained with a two-step procedure.
First, an optimally-smooth subspace is extracted 
from the Hilbert space (i.e., ``disentangled''), using a procedure
recently introduced \cite{Souza02} and specifically adapted here to the case of
$\Gamma$-sampling.
This $N$-dimensional subspace  
consists of a linear combination of all 
$N_o$ eigenstates at $\Gamma$ that fall inside a given energy window (with
$N_o > N$), and is chosen so that its dispersion across the 
BZ is as small as possible: this is akin to say that its
projection operator has a minimal change of character as the BZ is swept, 
or that the basis set obtained 
is as complete as possible. This subspace is identified by its projection operator
$\hat{P}^{({\bf \Gamma})}=\sum_{n=1}^N | \phi_n \rangle \langle \phi_n |$, where
$\vert \phi_n \rangle =\sum_{m=1}^{N_o} A_{mn} \vert \psi_m \rangle$.
The $N_o \times N$ matrix $A$  (such as $A^\dagger A={\bf I}$)
is the key quantity to be determined, and is chosen so that 
the projection operator at $\Gamma$ has maximal overlap 
with its first neighbors, i.e., it maximizes
$\sum_l W_l {\rm Tr}[ \hat{P}^{({\bf \Gamma})} \hat{P}^{({\bf G}_l)} ] $, where 
the neighbors ${\bf G}_l$ and the weight factors $W_l$ are chosen 
according to Refs. \cite{Marzari97,Silvestrelli98}.
Once this optimally-smooth manifold is extracted, a standard localization
procedure can be applied on the $\{\vert \phi_n \rangle\}$ to obtain MLWFs $\{\vert w_n \rangle\}$. 
\begin{figure}[tb]
\centering
\includegraphics[width=2.5in]{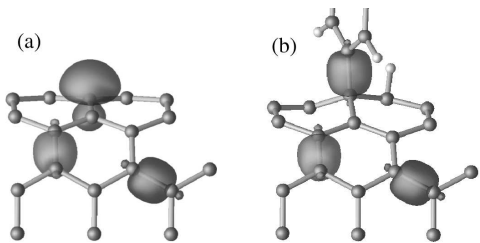}
\caption{MLWFs obtained from the disentanglement and localization
procedure. We show in
(a) the 3 inequivalent MLWFs obtained for a pristine metallic (5,5) CNT, 
clearly corresponding to a chemical picture of $p_z$ and $sp^2$ orbitals, and in
(b) the same MLWFs for the case when an aminophenyl group (1\% coverage) has
been covalently attached.} 
\label{fig:wan_ac5}
\end{figure}
\\
{\it Band Structure and Quantum Conductance:}  
The procedure outlined above corresponds to an
inverse mapping of first-principles, extended Bloch orbitals 
into tight-binding orbitals
$\vert w_{n{\bf R}}\rangle = \vert w_n ( {\bf r}-{\bf R})\rangle$
that are then periodically repeated to generate, via
Bloch sums, wavefunctions at any arbitrary {\bf k}:
$\vert\psi_{n{\bf k}}\rangle=\frac{1}{\sqrt {N_R}} \sum_{\bf R} e^{i{\bf k}\cdot
{\bf R}} \vert w_{n{\bf R}}\rangle$ ({\bf R} and {\bf k} refer to the direct
and reciprocal spaces of the supercell). 
Localization allows us to
neglect in the resulting Hamiltonian matrix 
all terms $\langle w_{m{\bf R}}\vert\hat{H}\vert w_{n{\bf R^\prime}}\rangle$
for which $\vert {\bf r}_{m{\bf R}}-{\bf r}_{n{\bf R^\prime}} \vert$$>$$L$, with
${\bf r}_{m{\bf R}}$ the center of $\vert w_{m{\bf R}} \rangle$, and
$L$ a cutoff distance 
determined by the spread of the MLWFs.
Since these MLWFs
span both the occupied and unoccupied subspaces, 
they will be well-localized and decay exponentially 
\cite{He01} even in metals.
The Hamiltonian matrix becomes diagonally-dominant, and is straightforwardly given by
\begin{equation}
\langle\psi_{m{\bf  k}} \vert  {\hat H}  \vert \psi_{n{\bf  k}}\rangle =
\langle w_{m{\bf 0}} | \hat{H} | w_{n{\bf 0}} \rangle
+\sum_{{\bf R}} e^{i{\bf  k}\cdot {\bf R}}  
\langle w_{m{\bf 0}} | \hat{H} | w_{n{\bf R}} \rangle
\label{eq:tbhamil}
\end{equation}
where the sum runs only over the very few supercells (two, for 1-dimensional 
nanotubes) interacting with the 
one at the origin. 
The advanced and retarded Green functions $G^{a,r}_C$ of
the conductor and the couplings with the left or right lead
$\Gamma_{L,R}$  can be derived from the matrix elements of
the Hamiltonian,
providing the equilibrium Landauer conductance 
${\cal  G}(E)=\frac{2  e^2}{h}{\rm Tr}(\Gamma_L G^r_C \Gamma_R G^a_C)$ 
(see Refs. \cite{Calzolari04,Datta,Nardelli99,Taylor01} for detailed explanations
of this approach and its implementation in periodic boundary conditions).
As a validation, we first calculate the band structure and the quantum conductance of a
pristine (5,5) CNT in a 100-atom supercell. 
The manifold of interest is that
spanned by the bonding combination of $sp^2$ orbitals 
(the graphitic backbone) and by the $p_z$ orbitals 
($\pi$ and $\pi^\ast$ bands), so the
disentanglement and localization
procedure uses a target dimension of 2.5 orbitals per atom.
We show in Fig. \ref{fig:wan_ac5} the resulting MLWFs,
confirming this intuitive chemical picture. In addition, the right panel shows the case
of an aminophenyl functionalization, where the ``half-filled'' $p_z$ orbital is replaced
by an $sp^3$ bonding orbital. The two other MLWFs barely show any
change, clearly hinting at the fact that maximal-localization corresponds also to
maximal-transferability of the MLWFs building blocks.
\begin{figure}[t]
\centering \includegraphics[width=2.8 in]{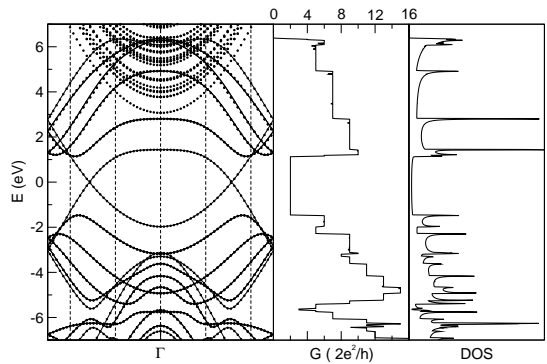}
\caption{Band structure, quantum conductance, and density of states of a (5,5) CNT as calculated in the
present approach based on MLWFs (solid lines), and compared with the results of a full diagonalization (circles). The
band structure has been unfolded over the BZ corresponding to the primitive 20-atom unit cell; the five vertical dashed
lines indicate the 5 {\bf k}-points corresponding to $\Gamma$ in a 100-atom supercell.}
\label{fig:bandtmdos}
\end{figure}
Once the (short-ranged) matrix elements of the Hamiltonian have been determined,
the band structure at any point in the BZ can be obtained with negligible computational costs by diagonalization of
Eq. (\ref{eq:tbhamil}).
Results are shown in 
Fig. \ref{fig:bandtmdos}, and compared 
with the band structure obtained from a diagonalization in a complete
plane-wave basis. 
The agreement between the two calculations is excellent - essentially a proof
of the proper disentanglement of the two manifolds (represented by solid lines and circles)
and localization of the MLWFs.
The extrema in the band structure and the steps and peaks in
the quantum conductance and the density of states are also all in perfect agreement. 
Even more significantly, the massless bands, the quantum conductance, and the
density of states around the Fermi energy 
clearly display the metallic character of the nanotube, 
{\it notwithstanding the fact that the
MLWFs have been obtained from the eigenstates at $\Gamma$ 
in the presence of a pseudo-gap of $\sim$2 eV in the folded BZ}.
The reason for this agreement is that,
moving away from $\Gamma$, the change in character of the occupied manifold
is described exactly, for all practical purposes, by the 
unoccupied part at $\Gamma$ of our extracted manifold. 
\begin{figure}[b]
\includegraphics[width=2.8 in]{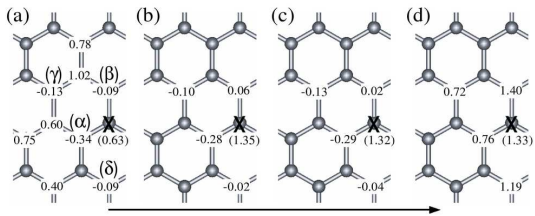}
\caption{
$\Delta E_2$ in eV 
as a function of the position of the functional group FG$_2$ when
the group
FG$_1$ is attached in the position marked by {\sf X} 
($\Delta E_1$ is indicated in parenthesis)
on a (5,5) CNT. 
FG$_1$ and FG$_2$ are, respectively,
(a) H and H, (b) nitrophenyl and H, (c) aminophenyl and H, (d) phenyl and phenyl
(in our calculation, the nitro and amino residues are in the para position).
The arrow lies parallel to the nanotube axis, and we denoted with $\alpha$, $\beta$, 
$\gamma$, and $\delta$ the arrangements for four of the most stable pairings.}
\label{fig:pattern} 
\end{figure}
\\
These tools now allow us to study with full first-principles accuracy
very complex structures, where the MLWFs obtained from a calculation at $\Gamma$
or even from an isolated fragment (e.g., a hydrogenated nanotube fragment) will reproduce 
accurately the fine details of a much larger extended system. We focus here on 
metallic CNTs whose sidewalls are functionalized for tens of nanometers 
with a disperse array of covalent ligands.
Functionalization with aryl groups has been recently accomplished \cite{Tour01}, and
our goal is to characterize the effects 
of electronegative or electropositive ligands such as
nitro- or amino-phenyls
on the band structure and 
transport properties of CNTs. 
It is important to stress that the addition of a covalent ligand can
affect nearby sidewall carbons, and expedite further reactions 
that reduce the local strain energy or minimize
the disruption of the conjugated manifold, as observed in
fullerenes \cite{Fullerene}. 
While exhaustive studies are necessary to fully characterize
the thermodynamic stability of the functional group decorations \cite{Stojkovic03,Kelly, Worsley}
or the subtle spin polarization they might induce \cite{Duplock04},
we found that, once a functional group (FG$_1$) is covalently bound to the 
sidewall, the attachment of a second one (FG$_2$) to a nearby carbon
becomes more facile.
These reactions are strongly site-dependent;
we present in Fig. \ref{fig:pattern} the first and second attachment energies,
$\Delta E_1$ and $\Delta E_2$, corresponding to 
${\rm CNT}+{\rm (FG_1)H}\rightarrow{\rm CNT(FG_1)}+{\rm 1/2 H_2}$
and 
${\rm CNT(FG_1)}+{\rm (FG_2)H}\rightarrow{\rm CNT(FG_1)(FG_2)}+{\rm 1/2 H_2}$  
respectively \cite{RXN}.
\begin{figure}[t]
\centering \includegraphics[width=2.8 in]{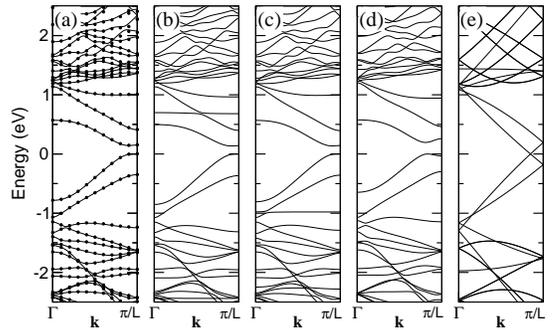}
\caption
{
Band structure of a (5,5) CNT decorated 
with a periodic array of functionalization pairs 
in position $\alpha$ (1 pair every 100 carbons):
(a) H pair
(dots are from a Quantum-Espresso calculation \cite{ESPRESSO}, for comparison),
(b) nitrophenyl and H, (c) aminophenyl and H,
(d) model calculation, where the 2 $p_z$ MLWFs on 
the functionalized sidewall carbons have been removed from the calculation,
(e) pristine nanotube.  }
\label{fig:funcband}
\end{figure}
In order to understand how
the electronic structure of the tube is  modified by
the presence of the ligands,
we first studied a periodic array of functionalized
pairs in the configuration ($\alpha$) of Fig. \ref{fig:pattern}, 
at a density of 1 pair per 100 carbons. 
The resulting band structures are shown in Fig. \ref{fig:funcband},
for the cases where the functional group pairs are composed either by two hydrogen atoms, 
or a nitro- or aminophenyl and a hydrogen. 
For comparison, we also report
the band structure of the pristine CNT (folded in the BZ of the
supercell), and the one obtained from the pristine CNT but removing the two $p_z$
MLWFs sitting on the two sidewall carbon atoms at ($\alpha$). 
Interestingly, these band structures are largely insensitive to the chemical
nature of the attached groups,
notwithstanding the charge
transfer taking place when  
e.g., aminophenyl is substituted for nitrophenyl
(0.1 $|e|$ per 100 atoms),
or the chemical differences
between a phenyl moiety and a hydrogen. 
The foremost effects on the
bands around the Fermi energy are described by a simple model in which the
functionalized sidewall carbons change hybridization from $sp^2$ to $sp^3$.
Functionalization transforms the $p_z$ MLWF into a bonding $sp^3$ MLWF whose
on-site energy, roughly lowered by 7 eV, 
removes it altogether from the $\pi$ conduction manifold;
this effect is clearly captured by the model
(Fig. \ref{fig:funcband}(d)).
A perfectly periodic array of functional pairs might be quite
unrealistic; from the transport point of view, the fundamental
quantity of interest is the reduction in the transmission probability as
more and more groups are attached to a 
functionalized segment
of an otherwise pristine nanotube (see the top of Fig. \ref{fig:scatter}). 
Our approach is particularly suited
to study such complex cases.
We show first in the left panels of Fig. \ref{fig:scatter}
the Landauer conductance as a function of energy for an
infinite (5,5) CNT with either 
one pair attached
in positions ($\alpha$), ($\beta$), ($\gamma$), and ($\delta$) or a single
covalent functional group.
The covalently-bonded groups act as strong
scattering centers, reducing the conductance at the Fermi energy by
19\%, 8\%, 21\%, 17\% and 42\% respectively. 
Again, the dominant factors are geometric and 
similar results are obtained, say, from hydrogen attachments (dashed line) or
the removal of $p_z$ MLWFs on the relevant sidewall carbons (solid line).
Once a random array of functional groups covers the nanotube for tens
of nanometers - comparable to the lengths of pristine tubes for which
we can observe ballistic transport \cite{Javey04} -
the conductance of the tube drops dramatically, as shown in the right panels of
Fig. \ref{fig:scatter}, for the case of a functionalized conductor 12.3 or 37.0 nm 
long (corresponding to 1000 or 3000 carbons) between semi-infinite leads.
We note that substitutional dopants such as boron and nitrogen 
also introduce substantial (albeit weaker) disorder, and can
change the transport regime from ballistic to localized \cite{Latil04}.
Among the patterns studied, 
$\beta$ turns out to be the weakest scattering center, 
since it is a paired defect that preserves the mirror plane containing the tube axis
and is almost transparent to either of the two Bloch states
with well-defined parity over the mirror plane
(a similar result has been discussed for
symmetric Stone-Wales defects on a (10,10) CNT \cite{Choi99}).
\\
In conclusion, we  have  presented a  novel  approach to  calculate
the band structure and transport properties  of  complex nanostructures
from large-scale electronic-structure calculations at the $\Gamma$ point
via an essentially exact inverse mapping 
into explicit and localized tight-binding orbitals (MLWFs).
These MLWFs act as electronic-structure building blocks, 
representing at the same time 
a minimal and most accurate basis set for the Hilbert space at hand
and providing a natural avenue to map first-principles calculations
into model Hamiltonians. 
An application of this approach
to covalently-functionalized metallic CNTs provides
a clear correlation between 
chemical bonding and transport properties, showing that
the band structure around the Fermi
energy is strongly influenced by  
the $sp^3$ functionalization pattern disrupting the
conjugation  network.
Common aryl functionalizations were found to be more stable
when clustered with saturating hydrogens; even when paired, they still
act as strong scattering centers that degrade the ballistic conductance
of carbon nanotubes already at low degrees of functionalization.
\\
The authors  would like to thank Y. Wu (Harvard University),
P. Giannozzi (Scuola Normale Superiore), and M. Sharma and R. Car (Princeton 
University) for the collaborative development of an ultrasoft implementation 
of MLWFs in the Quantum-Espresso package.  This research has been
supported by MIT Institute for Soldier Nanotechnologies 
(ISN-ARO DAAD 19-02-D-0002) and the National Science Foundation
(NSF-NIRT DMR-0304019).
M.B.N. acknowledges funding from the Department of Energy
(DE-AC05-00OR22725).

\end{document}